\documentclass[a4paper,prl,twocolumn,superscriptaddress,showpacs]{revtex4}

\pdfoutput=1
\usepackage{amssymb}
\usepackage{amsmath}
\usepackage{epsfig}
\usepackage{color}
\usepackage{graphics, graphicx}
\usepackage{bbold}
\usepackage{psfrag}
\usepackage{mathcomp}
\usepackage{subfigure}
\usepackage{verbatim}
\usepackage{color}
\usepackage[colorlinks,citecolor=blue]{hyperref}

\begin{document}

\date{\today}
\pacs{03.75.Ss, 03.75.Lm, 05.30.Fk}
\title{Topological Fulde-Ferrel-Larkin-Ovchinnikov states in Spin-orbit Coupled Fermi Gases}
\author{Wei Zhang}
\email{wzhangl@ruc.edu.cn}
\affiliation{Department of Physics, Renmin University of China, Beijing 100872, People's Republic of China}
\author{Wei Yi}
\email{wyiz@ustc.edu.cn}
\affiliation{Key Laboratory of Quantum Information, University of Science and Technology of China,
CAS, Hefei, Anhui, 230026, People's Republic of China}

\begin{abstract}
Pairing in an attractively interacting two-component Fermi gas in the absence of the inversion symmetry and/or the time-reversal symmetry may give rise to exotic superfluid states. Notable examples range from the Fulde-Ferrell-Larkin-Ovchinnikov (FFLO) state with a finite center-of-mass momentum in a polarized Fermi gas, to the topological superfluid state in a two-dimensional Fermi gas under Rashba spin-orbit coupling and an out-of-plane Zeeman field. Here, we show that a topological FFLO state can be stabilized in a two-dimensional Fermi gas with Rashba spin-orbit coupling and both in-plane and out-of-plane Zeeman fields. We characterize the topological FFLO state by a non-trivial Berry phase, and demonstrate the stability region of the state on the zero-temperature phase diagram. Given its unique properties in both the quasi-particle dispersion spectra and the momentum distribution, signatures of the topological FFLO state can be detected using existing experimental techniques.

\end{abstract}
\maketitle


\emph{Introduction}.--
Since its original proposal in 1960s, the search for the unconventional pairing states with finite center-of-mass
momentum has caught a considerable amount of attention in different physical contexts~\cite{fflo,casalbuoni-04}, e.g., heavy fermions~\cite{radovan-03}, dense quark matter~\cite{alford-01},
and ultracold atomic gases~\cite{liao-10}, etc.
Initially proposed as a compromise between superconductivity and finite magnetization, the key
ingredient of this so-called Fulde-Ferrel-Larkin-Ovchinnikov (FFLO) state is a pairing mechanism between
fermions having a finite center-of-mass momentum. In the weak coupling limit, this can be achieved by
pairing particles residing either on distinct Fermi surfaces, as in the case of spin-polarized systems where
spin-up and spin-down Fermi surfaces are mismatched, or on a single deformed Fermi surface which
breaks the spatial inversion symmetry. The latter possibility has been discussed in the context
of non-centrosymmetric superconductors, where the presence of Rashba spin-orbit coupling (SOC) and external magnetic
field would lead to a non-uniform superconducting state~\cite{agterberg-03, samokhin-04, kaur-05}.

The pairing physics in spin-orbit coupled Fermi systems is particularly interesting due to the lack of inversion 
symmetry. The study of exotic pairing superfluid states in these systems has attracted much attention recently, partly due to the
realization of synthetic spin-orbit coupling in ultracold atoms~\cite{lin-11,wang-12, cheuk-12, zhang-12}.
Theoretical investigation has demonstrated that the interplay of SOC, pairing superfluidity and effective Zeeman
fields can lead to exotic superfluid phases in various dimensions~\cite{zhang-08, sato-09, vyasanakere-11,
gong-11, yu-11, hu-11, iskin-11, yi-11, dellanna-11, gong-12, han-12, he-12, zhou-11, yang-12, yi-12,
han-12b, iskin-12, wu-13, zhou-13, zheng-13, dong-13, iskin-13, xu-13}. Notably, since the presence 
of SOC mixes different spin states, both intra- and inter-branch pairings can take place and the competition between 
them results in rich phase structures. An important example here is the topological superfluid state in a two-dimensional (2D)
Fermi gas with Rashba spin-orbit coupling and an out-of-plane Zeeman field. 
When the chemical potential lies within or below the gap opened by the out-of-plane Zeeman field, 
the subsequent intra-branch pairing results in a topological superfluid (TSF) state, in which a chiral 
Majorana edge mode is protected by the gap in the bulk. As both the inversion and the time-reversal symmetries are 
broken in the system, the topological superfluid state here belongs to class D.
\begin{figure}[tbp]
\includegraphics[width=8.0cm]{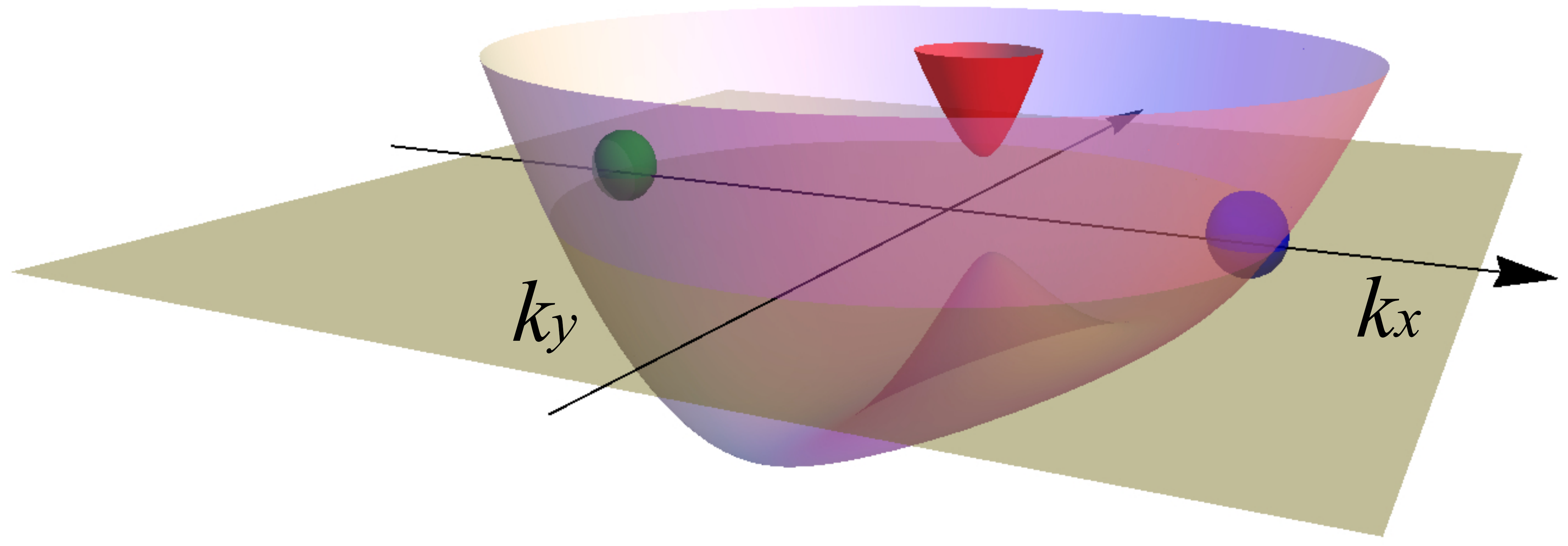}
\caption{(Color online) Illustration of pairing within the lower helicity branch, which can lead to 
a topological FF state in the presence of both out-of-plane and in-plane Zeeman fields.}
\label{fig:illustration}
\end{figure}

In this work, we show that a topological FFLO state can be stabilized in a two-dimensional Fermi gas with Rashba 
spin-orbit coupling and both in-plane and out-of-plane Zeeman fields. Similar to the case of a topological superfluid state, 
in the weak coupling limit, the emergence of the topological FFLO state can be understood as a result of single-band 
pairing within the lower helicity branch. As illustrated in Fig.~\ref{fig:illustration}, the application of the 
additional in-plane Zeeman field introduces deformation of the single-particle dispersion and, as a consequence, 
drives the system towards a more stable pairing state with a single-component non-zero center-of-mass momentum, 
i.e., the Fulde-Ferrel (FF) state. The resulting pairing state would preserve all topological properties provided that the 
deformation of the Fermi surface is not drastic enough to violate the single-band pairing scenario. This last condition 
is equivalent to the requirement that the introduction of the in-plane Zeeman field does not close the bulk gap. Thus, 
the topological nature of this state is protected by a full gap of quasi-particle spectra, and the topological FF (tFF) state 
belongs to the same classification as the topological superfluid state found in 2D Fermi gases with Rashba SOC~\cite{zhou-11}. 
We note that the center-of-mass momentum of this tFF state is antiparallel to the in-plane Zeeman field. 
By mapping out the zero-temperature phase diagram, we further discuss the competition between various FF states 
with different center-of-mass momentum. In particular, we find a nodal FF (nFF) state and characterize the evolution 
of its non-trivial gapless contours in momentum space. The tFF and nFF states should leave features in the spin-selective 
momentum distribution and momentum-resolved radio-frequency spectroscopy, respectively, which can in principle be 
detected using the existing experimental techniques.


\emph{Model}.--
We consider a two-component Fermi gas in two dimensions with a Rashba type SOC
and cross Zeeman fields, where the Hamiltonian can be written as
\begin{eqnarray}
H &=& \sum_{\mathbf{k},\sigma=\uparrow,\downarrow}\xi_{\bf k}a^{\dag}_{\mathbf{k}\sigma}a_{\mathbf{k}\sigma}
-h\sum_{\mathbf{k}} (a^{\dag}_{\mathbf{k}\uparrow}a_{\mathbf{k}\uparrow}
-a^{\dag}_{\mathbf{k}\downarrow}a_{\mathbf{k}\downarrow})
\nonumber \\
&& +\sum_{\mathbf{k}}\left\{
[\alpha (k_x+ik_y)-h_x] a^{\dag}_{\mathbf{k}\uparrow}a_{\mathbf{k}\downarrow}
+ \textrm{H.C.} \right\}
\nonumber\\
&& +U\sum_{\mathbf{k},\mathbf{k}',\mathbf{q}}a^{\dag}_{\mathbf{k}+\mathbf{q}\uparrow} a^{\dag}_{-\mathbf{k}+\mathbf{q}\downarrow}a_{-\mathbf{k}'+\mathbf{q}\downarrow}a_{\mathbf{k}'+\mathbf{q}\uparrow}.
\end{eqnarray}
Here, $\xi_{\bf k}=\epsilon_{\bf k}-\mu$, $\epsilon_{\bf k}=\hbar^2k^2/2m$, $a_{\mathbf{k}\sigma}$
($a^{\dag}_{\mathbf{k}\sigma}$) is the annihilation (creation) operator for the hyperfine spin state $\sigma$
with $\sigma=(\uparrow,\downarrow)$, $m$ is the atomic mass, $\alpha$ denotes the strength of the SOC,
and H.C. stands for Hermitian conjugate. The out-of-plane $(h)$ and in-plane $(h_x)$ Zeeman fields
can be effectively induced depending on how the synthetic SOC is implemented. As an example, $h$ and $h_x$
are proportional to the effective Rabi-frequency and the two-photon detuning, respectively, of the Raman process
in the current experimental scheme~\cite{wang-12, cheuk-12}. The bare $s$-wave interaction rate $U$ should be renormalized
as \cite{randeria-89}: $1/U=-{\cal S}^{-1} \sum_{\mathbf{k}}1/(E_b+2\epsilon_{\mathbf{k}})$, where ${\cal S}$ is
the quantization area, and $E_b$ is the binding energy of the two-body bound state in two dimensions without SOC,
which can be tuned, for instance, via the Feshbach resonance technique.

We focus on the zero-temperature properties of the Fulde-Ferrell (FF) pairing states with a single valued
center-of-mass momentum on the mean-field level \cite{fflo}. This should provide a qualitatively correct phase
diagram at zero temperature. The effective mean field Hamiltonian can then be arranged into a matrix form
in the hyperfine spin basis $\left\{a_{\mathbf{k}\uparrow},a^{\dag}_{\mathbf{Q}-\mathbf{k}\uparrow},
a^{\dag}_{\mathbf{Q}-\mathbf{k}\downarrow}, a_{\mathbf{k}\downarrow}\right\}^{T}$
\begin{eqnarray}
H_{\text{eff}}&=&\frac{1}{2}\sum_{\mathbf{k}}\begin{pmatrix}
\xi_{\bf k} - h & \Delta_{\bf Q} & 0 &  \Lambda_{\bf k} \\
 & - \xi_{{\bf Q}-{\bf k}} - h & -\Lambda_{{\bf Q}-{\bf k}} & 0 \\
 & & -\xi_{{\bf Q}-{\bf k}}+h & -\Delta_{\bf Q}^* \\
 & & & \xi_{\bf k} + h
\end{pmatrix}\nonumber\\
&& + \sum_{\mathbf{k}}\xi_{|\mathbf{Q}-\mathbf{k}|}-\frac{|\Delta_{\bf Q}|^2}{U},\label{eqnHeff}
\end{eqnarray}
where $\xi_{{\bf Q}-{\bf k}} = \epsilon_{{\bf Q} - {\bf k}} - \mu$, $\Lambda_{\bf k} = \alpha (k_x + ik_y) - h_x$,
and the order parameter $\Delta_{\bf Q}=U\sum_{\mathbf{k}}\left\langle a_{\mathbf{Q}-\mathbf{k}\downarrow}
a_{\mathbf{k}\uparrow} \right\rangle$. It is then straightforward to diagonalize the effective Hamiltonian
and evaluate the thermodynamic potential at zero temperature
\begin{equation}
\Omega=\sum_{\mathbf{k}}\xi_{|\mathbf{Q}-\mathbf{k}|}+\sum_{\mathbf{k},\nu}
\Theta(-E^{\eta}_{\mathbf{k},\nu})E^{\eta}_{\mathbf{k},\nu}-\frac{|\Delta_{\bf Q}|^2}{U},
\label{eqnOmega}
\end{equation}
where the quasi-particle ($\eta = +$) and quasi-hole ($\eta = -$) dispersions $E^{\eta}_{\mathbf{k},\nu}$
($\nu=1,2$) are the eigenvalues of the matrix in Hamiltonian (\ref{eqnHeff}), and $\Theta(x)$ is the Heaviside
step function. Without loss of generality, we assume $h,h_x>0$, $\Delta_0=\Delta$, and $\Delta_{\bf Q}$
to be real throughout the work. The pairing order parameter $\Delta_{\mathbf{Q}}$ as well as the center-of-mass
momentum $\mathbf{Q}$ for the pairs can then be found by minimizing the thermodynamic potential
in Eq. (\ref{eqnOmega}).
\begin{figure}[tbp]
\includegraphics[width=8.0cm]{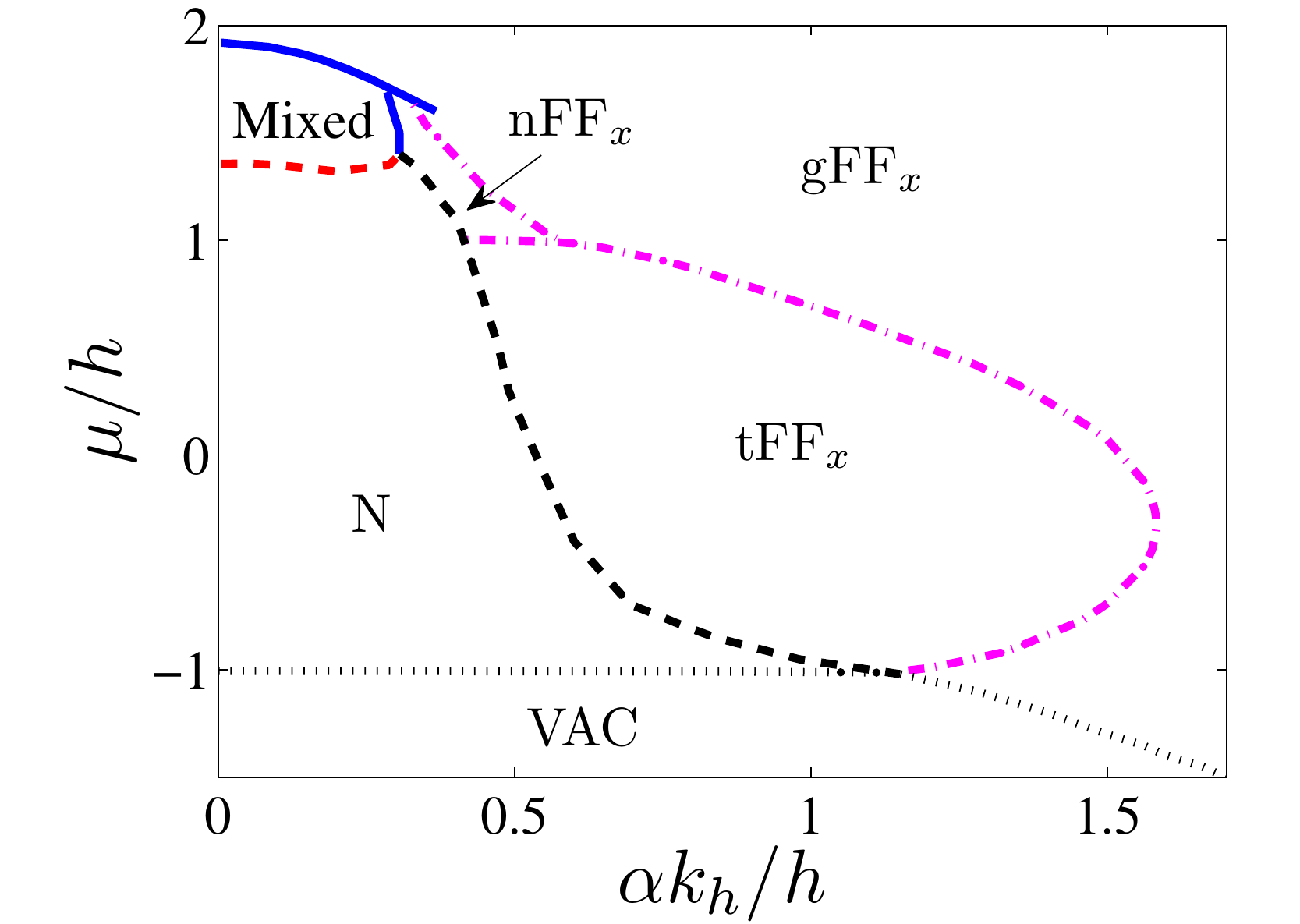}
\caption{(Color online) Phase diagrams on the $\mu$--$\alpha$ plane for $E_b/h=0.5$, $h_x/h=0.1$.
The solid curves are first-order boundaries, while the dashed-dotted curves represent phase boundaries of continuous phase transitions. The dashed curves surrounding the normal region (N) are the threshold with
$\Delta/h=10^{-3}$, while the dotted curves are the boundary against vacuum. The axial Zeeman field $h$ is taken to be the unit of energy, while the unit of momentum
$k_h$ is defined through $\hbar^2k_h^2/2m=h$. }
\label{fig:phasediag}
\end{figure}

In general, the Hamiltonian (\ref{eqnHeff}) cannot be diagonalized analytically, and the thermodynamic
potential needs to be evaluated numerically. However, for pairing states with zero center-of-mass momentum
($Q=0$), analytical form of the dispersion spectrum can be obtained for $h_x =0$~\cite{zhou-11}.
In this case, a fully gapped topological superfluid phase can be stabilized in a fairly large
parameter region. The topological nature of this phase is characterized by a non-trivial topological
number, and is protected by the underlying particle-hole symmetry. For the case of $h_x \neq 0$, it can be
proved that the zero center-of-mass momentum state becomes unstable against an FF state with
pairing momentum ${\bf Q} = Q_x {\hat x}$. Thus, a topologically non-trivial FF state can be expected
provided that the in-plane field $h_x$ is not large enough to close the gap of the bulk.
This topological FF state hence belongs to the same classification as the TSF phase, and
acquires all topological features including gapless edge modes and Majorana fermions in vortex cores.
\begin{figure}[tbp]
\includegraphics[width=8cm]{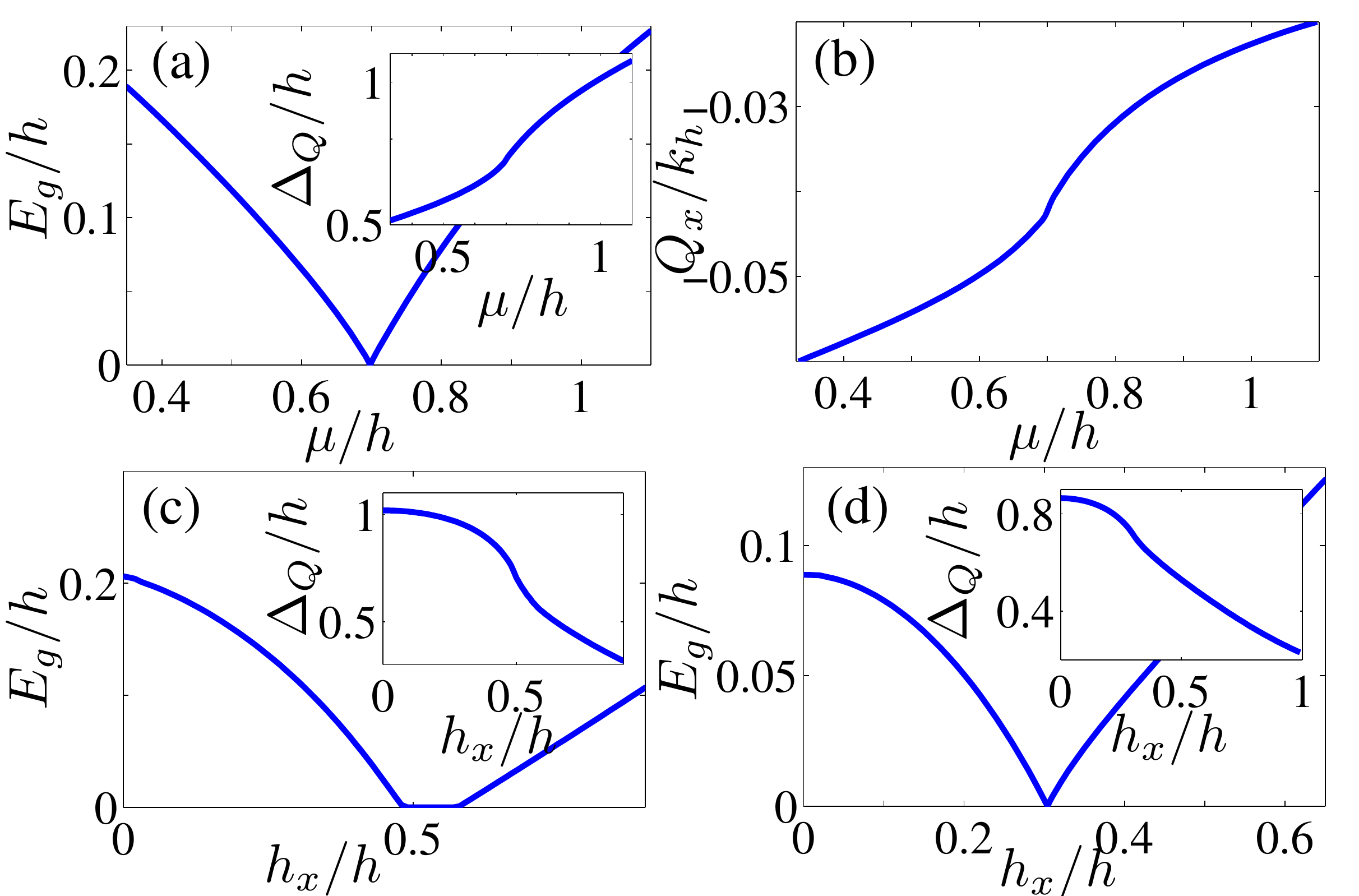}
\caption{(Color online) (a-b) Evolution of the minimum excitation gap (a), the order parameter (a:inset),
and pairing momentum (b) with increasing chemical potential. A full gap closes and opens again by
traversing from the tFF$_x$ state to the gFF$_x$ state by passing through a continuous phase boundary.
In these subplots, $h_x/h=0.1$.
(c-d) Evolution of the minimum excitation gap and the pairing order parameter (insets) with increasing
in-plane field $h_x$, with: (c) $\mu/h=1$; and (d) $\mu/h=0.8$. Other parameters
used in this figure are $E_b/h=0.5$, $\alpha k_h/h=1$.}
\label{fig:scaling}
\end{figure}
%


\emph{Phase diagram and topological FF state}.--
We map out the phase diagram on the $\mu$--$\alpha$ plane with fixed $h$ and $h_x \neq 0$ (see Fig. \ref{fig:phasediag}).
Under the local density approximation, the phases traversed by a downward vertical line in the diagram
represent those one should encounter by moving from a trap center to its edge.
From Fig. \ref{fig:phasediag}, we see that the topological superfluid phase in a 2D polarized Fermi gas
with Rashba SOC and zero in-plane field is now replaced by a topological FF phase
with center-of-mass momentum along the $x$-dirextion (tFF$_x$), as we have anticipated.
To characterize the non-trivial topological nature of this state, we further calculate the Berry phase
associated with each quasi-particle ($\eta=+$) or quasi-hole ($\eta=-$) bands
\begin{eqnarray}
\label{eqn:berryphase}
\gamma_{\nu=1,2}^{\eta}= \frac{1}{2\pi} \sum_{\nu} \int dk_x dk_y \Gamma_{\nu}^{\eta}(k_x,k_y),
\end{eqnarray}
where the Berry curvature is defined as~\cite{xiao-10}
\begin{eqnarray}
\label{eqn:berrycurvature}
\Gamma_{\nu}^\eta ({\bf k}) =
i \sum_{\ell \neq \ell^\prime}
\frac{\langle \ell \vert \partial_{k_x} H \vert \ell^\prime \rangle
\langle \ell^\prime \vert \partial_{k_y} H \vert \ell \rangle
- (k_x  \leftrightarrow k_y)}
{(E_{{\bf k}, \nu}^{\eta} -  E_{{\bf k}, \nu^\prime}^{\eta^\prime})^2}
\end{eqnarray}
with $\ell \equiv (\nu, \eta)$ the shorthand notation.
The Berry phase of the superfluid phase is then a summation over the contribution $\gamma_{\nu}^{\eta=-}$
from the two occupied quasi-hole bands. A numerical evaluation shows that the resulting Berry phase
vanishes in the topological trivial phase and becomes unity in the tFF$_x$ state. As we have discussed before,
the stabilization of the tFF$_x$ state is due to the SOC-induced single-branch pairing and the Fermi surface 
asymmetry.

The picture of single-branch pairing is complicated with increasing chemical potential or increasing
SOC intensity, such that particles on the higher helicity branch get involved into pairing.
Due to the spin mixing induced by the SOC, both intra- and inter-band pairings can take place
with center-of-momentum along either the $x$- or the $y$-direction. As a consequence, the ground
state of the system is the result of competitions between the various FFLO pairing states, as depicted
in Fig.~\ref{fig:phasediag}.

For strong SOC intensity, or equivalently weak out-of-plane Zeeman field,
the tFF$_x$ state is separated from a topologically
trivial FF state (depicted as gFF$_x$ in Fig.~\ref{fig:phasediag}) via a continuous phase transition.
This gFF$_x$ phase is also fully gapped and with center-of-mass momentum along
the $x$-axis. By tuning through the tFF-gFF phase boundary, the excitation gap
closes and opens again, while the pairing order parameter $\Delta_{Q}$ remains finite. This leads to a change of topology 
as the boundary is crossed.
A typical variation of the minimum excitation gap $E_g$, the pairing order parameter $\Delta_Q$,
and the pairing momentum $Q_x$ with increasing chemical potential are plotted in Figs.~\ref{fig:scaling}(a) 
and \ref{fig:scaling}(b), respectively.
\begin{figure}[tbp]
\includegraphics[width=8cm]{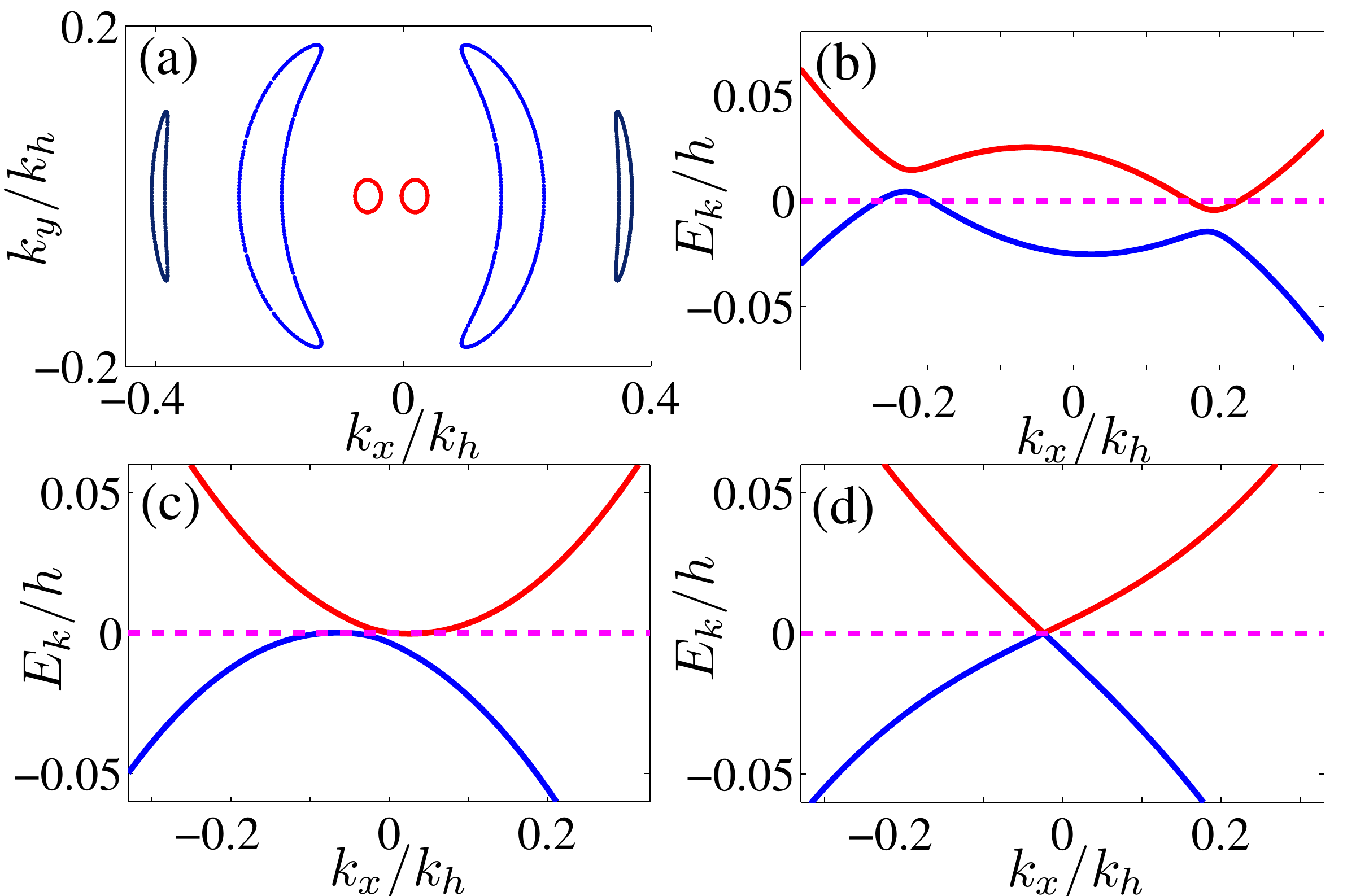}
\caption{(Color online)
(a) Evolution of gapless contours in the momentum space for the nFF$_x$ states with $\mu/h=1.002$ (red), 
1.05 (blue), and 1.15 (black). (b-d) Dispersion spectra of quasi-particle and quasi-hole along the $k_x$-axis, with
(b) $\alpha k_h/h=0.49$, $\mu/h=1.05$; (c) $\alpha k_h/h=0.49$, $\mu/h=1.002$;
and (d) $\alpha k_h/h=0.8$, $\mu/h=0.87$. At the topological phase boundaries, 
two gapless points exist between the nFF$_x$ and the tFF$_x$ states (c), and a single gapless 
point exists between the gFF$_x$ and the tFF$_x$ states.
Other parameters used in these plots are $E_b/h=0.5$ and $h_x/h=0.1$. }
\label{fig:dispersion}
\end{figure}

The inter-band pairing scenario becomes significant with decreasing SOC intensity or
increasing out-of-plane Zeeman field,
which can effectively polarize the helices within each helicity branch and hence hinder intra-brand
pairing. As a result, we identify a region where the globally stable state is a nodal FF phase
with pairing momentum along the $x$-direction, as denoted by nFF$_x$ in Fig.~\ref{fig:phasediag}.
This gapless superfluid state acquires two disconnected gapless contours in momentum
space, which shrink to two separated gapless points and disappear at the continuous phase boundaries.
A typical evolution of the two gapless contours are displayed in Fig.~\ref{fig:dispersion}(a) with increasing
chemical potential. Notice that at the phase boundaries between nFF$_x$ and the gapped FF states (gFF$_x$ or tFF$_x$),
the quasi-particle and quasi-hole dispersions touch the Fermi surface at different places [see Fig.~\ref{fig:dispersion}(c) for example]. 
This is in contrast to the phase boundary between the tFF$_x$ state and the gFF$_x$ state, where the gap closes 
at a single point $\mathbf{k}=(Q_x/2,0)$ in momentum space [see Fig.~\ref{fig:dispersion}(d)]. 
Hence, the phase boundary between the tFF$_x$ and the gFF$_x$ state can be worked out by examining the gap closing condition
\begin{equation}
h^2+h_x^2 - \left[ \Delta^2+\left(\frac{\hbar^2Q_x^2}{8m}-\mu\right)^2 \right]
-\alpha h_x Q_x+\alpha^2\frac{Q_x^2}{4}=0.
\end{equation}   
On the other hand, the phase boundary between the nFF$_x$ state and the tFF$_x$ state needs to be determined numerically.
Finally, we note that by further increasing the effective Zeeman field $h$, inter-band pairing with
center-of-mass momentum along other directions has to be taken into account, and the system
can be stabilized as a general FFLO state (labeled as {\it mixed} in Fig.~\ref{fig:phasediag}), where 
multiple FF states with various pairing momenta coexist~\cite{xu-13}.

We then investigate the effect of increasing in-plane Zeeman field $h_x$. In the weak coupling limit,
the presence of a larger $h_x$ lifts the upper helicity branch and enlarges the gap between two
branches. As a consequence, the tFF$_x$ state, which is dominated by the intra-band pairing
within the lower helicity branch, becomes stable in an extended parameter region, and the phase
boundaries surrounding the tFF$_x$ state are shifted towards larger values of chemical potential and
SOC intensity. In Fig.~\ref{fig:scaling}(c-d), we show the variation of the minimum gap and the pairing order
parameter with increasing in-plane field $h_x$, indicating the two representative evolution paths
of the system from gFF$_x$ to nFF$_x$ and eventually to tFF$_x$ [Fig.~\ref{fig:scaling}(c)], or
from gFF$_x$ directly to tFF$_x$ [Fig.~\ref{fig:scaling}(d)], depending on the starting point on the phase diagram.
\begin{figure}[tbp]
\includegraphics[width=8.6cm]{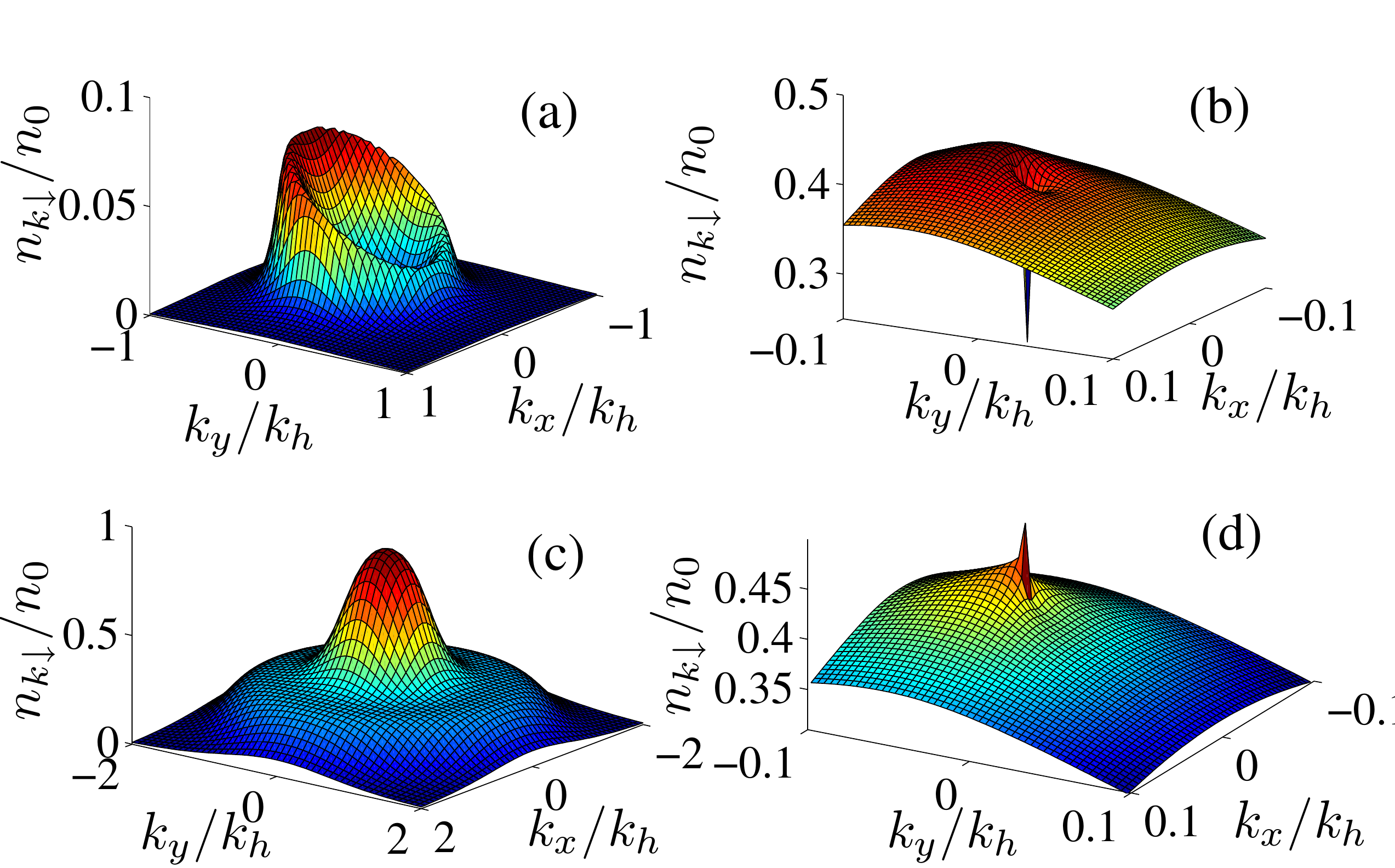}
\caption{(Color online) Density distribution in the momentum space for the minority component.
Parameters used in this plot are: $E_b / h = 0.5$, $h_x/h = 0.1$, $\alpha k_h / h =1$, and the chemical potentials
are: (a) $\mu/h = -0.85$, (b) $\mu/h=0.6977$,(c) $\mu/h = 1$, and (d) $\mu/h =0.6980$.}
\label{fig:ndk}
\end{figure}

\emph{Characterizing the FF states}.--
To characterize the properties of the various phases in the phase diagram, we calculate their respective
momentum distribution. In Fig.~\ref{fig:ndk}, we show the momentum profiles of the minority component
for cases within the tFF$_x$ region (a-b) and the gFF$_x$ region (c-d). It is apparent that the momentum
distribution in a topologically non-trivial phase is drastically different from that in a topological trivial phase.
In particular, the density profile in momentum space for the minority spin features a dip near zero momentum
in the tFF$_x$ phase, in contrast to a peak in the gFF$_x$ case. By extracting the momentum
distribution by species-selective time-of-flight imaging, this qualitative difference may serve as a
signature for the detection of topological FF state in the underlying system.

\emph{Conclusion}.--
In this manuscript, we investigate the pairing states of a two-dimensional Fermi gas with Rashba
spin-orbit coupling and both in-plane and out-of-plane Zeeman fields. We show that the BCS pairing
state becomes unstable towards an FFLO state with finite center-of-mass momentum as the
Fermi-surface becomes asymmetric in the presence of an in-plane Zeeman field.
In particular, we identify a topological FF state with center-of-mass momentum antiparallel to the
direction of the in-plane field. The topological nature of this tFF$_x$ phase is characterized by
a non-trivial Berry phase, which vanishes for a topological trivial state. We further map out the
zero-temperature phase diagram, where multiple FF states are separated by either first-order
or second-order phase transitions. These FF states are characterized by different quasi-particle
dispersion spectra and momentum distribution, which can be distinguished via spectroscopic
detection and species selective time-of-flight imaging technique, respectively.

\emph{Note added}.--
After finalizing the present manuscript, we notice a simultaneous work by Qu {\it et al.} who also discuss the existence of 
a topological FFLO state in Fermi systems with spin-orbit coupling~\cite{qu-13}.

\emph{Acknowledgements}.--
We thank Hong Yao and Ying Ran for helpful discussions. This work is supported by NFRP (2011CB921200, 2011CBA00200), NKBRP (2013CB922000), NNSF (60921091), NSFC (11105134, 11274009), the Fundamental Research Funds for the Central Universities (WK2470000006), and the Research Funds of Renmin University of China (10XNL016).


\begin{thebibliography}{99}

\bibitem{fflo} P. Fulde and R. A. Ferrell, Phys. Rev. {\bf 135}, A550 (1964); A. I. Larkin and Y. N. Ovchinnikov, Sov. Phys. JETP {\bf 20}, 762 (1965).

\bibitem{casalbuoni-04}
R. Casalbuoni and G. Nardulli, Rev. Mod. Phys. {\bf 76}, 263 (2004).

\bibitem{radovan-03}
H. A. Radovan, N. A. Fortune, T. P. Murphy, S. T. Hannahs, E. C. Palm, S. W. Tozer, and D. Hall,
Nature {\bf 425}, 52 (2003).

\bibitem{alford-01}
M. Alford, J. A. Bowers, and K. Rajagopal, Phys. Rev. D {\bf 63}, 074016 (2001).

\bibitem{liao-10}
Y. A. Liao, A. S. C. Rittner, T. Paprotta, W. Li, G. B. Partridge, R. G. Hulet, S. K. Baur, E. J. Mueller,
Nature {\bf 467}, 567 (2010).

\bibitem{agterberg-03} D. F. Agterberg, Physica C {\bf 387}, 13 (2003).

\bibitem{samokhin-04} K. V. Samokhin, Phys. Rev. B {\bf 70}, 104521 (2004).

\bibitem{kaur-05} R. P. Kaur, D. F. Agterberg, and M. Sigrist, Phys. Rev. Lett. {\bf 94}, 137002 (2005).

\bibitem{lin-11} Y.-J. Lin, K. Jim\'{e}nez-Garc\'{i}a, and I. B. Spielman, Nature (London) {\bf 471}, 83 (2011).

\bibitem{wang-12} P. Wang, Z.-Q. Yu, Z. Fu, J. Miao, L. Huang, S. Chai, H. Zhai, and J. Zhang,
Phys. Rev. Lett. {\bf 109}, 095301 (2012).

\bibitem{cheuk-12} L. W. Cheuk, A. T. Sommer, Z. Hadzibabic, T. Yefsah, W. S. Bakr, and M. W. Zwierlein,
Phys. Rev. Lett. {\bf 109}, 095302 (2012).

\bibitem{zhang-12} J.-Y. Zhang, {\it et al.}, Phys. Rev. Lett. {\bf 109}, 115301 (2012).

\bibitem{zhang-08} C. Zhang, S. Tewari, R. M. Lutchyn, and S. Das Sarma, Phys. Rev. Lett. {\bf 101}, 160401 (2008).

\bibitem{sato-09} M. Sato, Y. Takahashi, and S. Fujimoto, Phys. Rev. Lett. {\bf 103}, 020401 (2009).

\bibitem{vyasanakere-11} J. P. Vyasanakere, S. Zhang, and V. B. Shenoy, Phys. Rev. B {\bf 84}, 014512 (2011).

\bibitem{gong-11} M. Gong, S. Tewari, and C. Zhang, Phys. Rev. Lett. {\bf 107}, 195303 (2011).

\bibitem{yu-11} Z.-Q. Yu and H. Zhai, Phys. Rev. Lett. {\bf 107}, 195305 (2011).

\bibitem{hu-11} H. Hu, L. Jiang, X.-J. Liu, and H. Pu, Phys. Rev. Lett. {\bf 107}, 195304 (2011).

\bibitem{iskin-11} M. Iskin and A. L. Subasi, Phys. Rev. Lett. {\bf 107}, 050402 (2011).

\bibitem{yi-11} W. Yi and G.-C. Guo, Phys. Rev. A {\bf 84}, 031608(R) (2011).

\bibitem{dellanna-11} L. Dell'Anna, G. Mazzarella, and L. Salasnich, Phys. Rev. A {\bf 84}, 033633 (2011).

\bibitem{gong-12} M. Gong, G. Chen, S. Jia, and C. Zhang, Phys. Rev. Lett. 109, 105302 (2012).

\bibitem{han-12} L. Han and C. A. R. S\'{a} de Melo, Phys. Rev. A {\bf 85} 011606(R) (2012).

\bibitem{he-12} L. He and X.-G. Huang, Phys. Rev. Lett. {\bf 108}, 145302 (2012).

\bibitem{zhou-11} J. Zhou, W. Zhang, W. Yi, Phys. Rev. A {\bf 84}, 063603 (2011).

\bibitem{yang-12} X. Yang and S. Wan, Phys. Rev. A {\bf 85}, 023633 (2012).

\bibitem{yi-12} W. Yi and W. Zhang, Phys. Rev. Lett. {\bf 109}, 140402 (2012).

\bibitem{han-12b} L. Han and C. A. R. S\'{a} de Melo, arXiv:1206.4984.

\bibitem{iskin-12} M. Iskin and A. L. Subasi, arXiv:1211.4020.

\bibitem{wu-13} F. Wu, G.-C. Guo, W. Zhang, W. Yi, Phys. Rev. Lett. {\bf 110}, 110401 (2013).

\bibitem{zhou-13} X.-F. Zhou, G.-C. Guo, W. Zhang, W. Yi, Phys. Rev. A {\bf 87}, 063606 (2013).

\bibitem{zheng-13} Z. Zheng, M. Gong, X. Zou, C. Zhang, and G.-C. Guo,
Phys. Rev. A {\bf 87}, 031602(R) (2013).

\bibitem{dong-13} L. Dong, L. Jiang, and H. Pu, arXiv:1302.1189.

\bibitem{iskin-13} M. Iskin, arXiv:1304.1473.

\bibitem{xu-13} Y. Xu, C. Qu, M. Gong, and C. Zhang, arXiv:1305.2152.

\bibitem{conduit-08} G. J. Conduit, P. H. Conlon, and B. D. Simons, Phys. Rev. A {\bf 77}, 053617 (2008).


\bibitem{randeria-89} M. Randeria, J.-M. Duan, and L.-Y. Shieh, Phys. Rev. Lett. {\bf 62}, 981 (1989).

\bibitem{xiao-10} D. Xiao, M. C. Chang, Q. Niu, Rev. Mod. Phys. {\bf 82}, 1959 (2010).

\bibitem{qu-13} C. Qu, Z. Zheng, M. Gong, Y. Xu, L. Mao, X. Zou, G.-C. Guo, and C. Zhang, arXiv:1307.1207.

\end{thebibliography}
\end{document}